\documentclass[seceqn]{elsarticle}

\usepackage{graphicx,color,slashed,amsfonts,amsmath,amssymb,mathrsfs}

\newcommand{\sherpa}{{\sc Sher\-pa}}

\newcommand{\herwig}{{\sc Herwig}}
\newcommand{\madgraph}{{\sc Mad\-Graph}}
\newcommand{\madevent}{{\sc Mad\-E\-vent}}
\newcommand{\mgamc}{{\sc Mad\-Graph5\-\_aMC\-@\-NLO}}
\newcommand{\maddecay}{{\sc MadWidth}}

\newcommand{\feynrules}{{\sc FeynRules}}
\newcommand{\mathematica}{{\sc Mathematica}}
\newcommand{\feynarts}{{\sc FeynArts}}
\newcommand{\formcalc}{{\sc FormCalc}}

\newcommand{\oomega}{{\sc O'Mega}}
\newcommand{\calchep}{{\sc CalcHep}}
\newcommand{\comphep}{{\sc Comp\-Hep}}

\newcommand{\eg}[0]{{\it e.g.}}
\newcommand{\ie}[0]{{\it i.e.}}
\newcommand{\etc}[0]{\emph{etc.{}} }
\newcommand{\python}{{\sc Python}}

\begin{document}

\begin{frontmatter}

\title{Computing decay rates for new physics theories with \feynrules\ and \mgamc}

\author[a]{Johan Alwall}
\author[b]{Claude Duhr\corref{cor1}}
\author[c,d]{Benjamin Fuks}
\author[e]{Olivier Mattelaer}
\author[f]{Deniz Gizem \"Ozt\"urk}
\author[a,g]{Chia-Hsien Shen}

\address[a]{Department of Physics, National Taiwan University, 
   Taipei 10617, Taiwan}
\address[b]{Institute for Particle Physics Phenomenology, University of Durham,\\ 
    Durham, DH1 3LE, United Kingdom} 
\address[c]{Theory Division, Physics Department, CERN, CH-1211 Geneva 23, Switzerland}
\address[d]{Institut Pluridisciplinaire Hubert Curien/D\'epartement Recherches
    Subatomiques, Universit\'e de Strasbourg/CNRS-IN2P3, 23 Rue du Loess, F-67037
    Strasbourg, France}
\address[e]{Center for Cosmology, Particle Physics and Phenomenology, Universit\'e Catholique de Louvain, chemin du Cyclotron,2, 1347 Louvain-La-Neuve}
\address[f]{Institut f\"ur Theoretische Physik, Universit\"at Z\"urich, CH-8057 Z\"urich, Switzerland}
\address[g]{California Institute of Technology, Pasadena, CA 91125, USA}

\cortext[cor1]{Corresponding Author: Claude Duhr; Email: claude.duhr@cern.ch; Phone: +41 (0)22 767 2447}

\begin{abstract}
We present new features of the \feynrules\ and \mgamc\ programs for the
automatic computation of decay widths that consistently include channels
of arbitrary final-state multiplicity. The implementations are generic enough so that they can be used
in the framework of any quantum field theory, possibly
including higher-dimensional operators. We extend at the same time the conventions
of the Universal \feynrules\ Output (or UFO)
format to include decay tables and information on the total widths. We finally provide a set of representative examples of the usage of the new functions
of the different codes in the framework of the Standard Model, the Higgs Effective Field Theory,
the Strongly Interacting Light Higgs model and the Minimal Supersymmetric Standard Model and compare
the results to available literature and programs for validation purposes.


\end{abstract}

\end{frontmatter}

\newpage

\noindent {\bf PROGRAM SUMMARY}                                               \\
  \begin{small}
  {\bf Manuscript Title:}Computing decay rates for new physics theories with \feynrules\ and \mgamc          \\
  {\bf Authors:} Johan Alwall, Claude Duhr, Benjamin Fuks, Olivier Mattelaer, Deniz Gizem \"Ozt\"urk, Chia-Hsien Shen                                     \\
  {\bf Program Title:}   \maddecay                                   \\
  {\bf Journal Reference:}                                                    \\
  {\bf Catalogue identifier:}                                                 \\
  {\bf Licensing provisions:} None.                                           \\
  {\bf Programming language:} \mathematica\ \& \python.                                   \\
  {\bf Computer:} Platforms on which \mathematica\ and \python\ are available.                \\
  {\bf Operating system:} Operating systems on which \mathematica\ and \python\ are available.
  {\bf Classification:} 11.1 General, High Energy Physics and Computing.      \\
  \phantom{{\bf Classification:}} 11.6 Phenomenological and Empirical Models
                                 and Theories.                                \\
  {\bf External routines/libraries:} \feynrules\ 2.0 or higher.\\
  \mgamc\ 2.2 or higher.                              \\
  {\bf Nature of problem:} The program is a module for the \feynrules\ and \mgamc\ packages that allows the computation of tree-level decay widths for arbitrary new physics models. The module consists of two parts:
  \begin{enumerate}
  \item A \feynrules\ part, which allows one to compute analytically all tree-level two-body decay rates and to output them in the UFO format.                                                       
  \item A \mgamc\ part, which allows the numerical computation of many-body decay rates.
  \end{enumerate}
  {\bf Solution method:}
  \begin{enumerate}
  \item For the \feynrules\ part, the analytic expressions for the three-point vertices can be squared to obtain analytic formulas for two-body decay rates. 
  \item For the \mgamc\ part, \madgraph\ is used to generate all Feynman diagrams contributing to the decay, and diagrams that correspond to cascade decays are removed.
  \end{enumerate}
  {\bf Restrictions:} \mathematica\ version 7 or higher. As the package is a module relying on \feynrules\ and \mgamc\, all restrictions of these packages apply.                                                    \\
  {\bf Unusual features:} None.                                  \\
  {\bf Running time:} The computation of the Feynman rules from a Lagrangian, as well as the computation of the decay rates, varies with the complexity of the model, and runs from a few seconds to 
    several minutes. See Section \ref{sec:validation} of the present manuscript for 
    more information.\\
\end{small}
\newpage

\section{Introduction}
The Monte Carlo simulation of new physics models with new massive unstable particles
requires, to be practically useful, the calculation of the total and partial decay
widths for all particles. 
The higher the number of allowed decay channels, the more daunting it is to do this
by hand. Furthermore, depending on the mass hierarchy and
interactions among the particles, the computation of two-body decay rates might be 
insufficient, as higher-multiplicity decays might be the dominant decay modes for some
of the particles. Finally, the decay channels that are kinematically allowed are highly dependent 
on the mass spectrum of the model, so that the decay rates need to be reevaluated for every
 choice of the input parameters.
As a consequence, the computation of all the partial widths of all the particles that appear in
a model can be a complex task already at leading order.
For this reason, several tools dedicated to the computation of decay rates
in the context of specific beyond the Standard Model theories have been
developed in the past~\cite{Djouadi:1997yw,Heinemeyer:1998yj,Allanach:2001kg,Muhlleitner:2003vg,Porod:2003um,%
Djouadi:2006bz,Meade:2007js,SMcalc,Eriksson:2009ws,Frisch:2010gw,Das:2011dg,Hlucha:2011yk,Contino:2013kra,Baglio:2013iia}, while
many Monte-Carlo event generators are also able to compute partial widths
on their own via a dedicated phase-space integration \cite{Pukhov:2004ca,Belyaev:2012qa,Bahr:2008pv,Maltoni:2002qb,Alwall:2011uj,Gleisberg:2003xi,Kilian:2007gr,Hahn:2006ig,Klasen:2002xi}.
The aim of this paper is to present a way to compute automatically, using
the \feynrules~\cite{Christensen:2008py,Alloul:2013bka} and 
\mgamc~\cite{Alwall:2011uj,amcatnlo} frameworks, all the partial decay widths
for large classes of new physics models, in particular when
specific width calculators are not publicly available.

\feynrules\ is a \mathematica\ package that allows one to compute
the interaction vertices of any high-energy physics model
directly 
from its Lagrangian. The package contains a set of translation interfaces to
export the Feynman rules to various matrix 
element generators. The latter often require the widths of all the particles 
appearing inside the model to be given explicitly and to be defined as numerical input parameters.
The workflow to obtain an implementation of a new physics model that can be used for 
phenomenological studies has so far been the following\footnote{For the sake of generality, we ignore here external width
calculators.}:
\begin{enumerate}
\item obtain an implementation of the model into the matrix element generator of choice (for which an interface to \feynrules\ exists) where all the widths are set to some default values.
\item for each choice of the numerical input parameters, run the matrix element generator to compute the numerical values of the widths, and insert them back into the model implementation.
\end{enumerate}

While straightforward, it is clear that this process contains a lot of redundant workload. 
In particular, in many situations the dominant kinematically allowed decay channels are the 
two-body decays, which can easily be computed by squaring three-point vertices and 
multiplying by the appropriate phase-space factors. The first main technical advance of this paper is 
that we present an extension of the \feynrules\ package that allows one to compute analytically all
two-body decay rates, and thus to include their values into the output of the translation 
interfaces. Moreover, we have extended the Universal \feynrules\ Output
format (UFO)~\cite{Degrande:2011ua}, which is the standard
interface from \feynrules\ to \mgamc, {\sc GoSam}~\cite{Cullen:2011ac} and
\herwig++~\cite{Bahr:2008pv,Bellm:2013lba},
to include all these analytic formulas so that 
they can be used dynamically when generating a scattering process.

Two-body decays might however be insufficient and three (or even more)-body decays might be required for a reliable estimation of the total widths
of some of the particles.
However, computing all the analytic formulae for an arbitrary mass spectrum is beyond reach. In addition, one has to deal with the double counting coming from an intermediate propagator going
on-shell, which corresponds to a cascade of two-body decays. Therefore, the second main technical achievement of this paper is to introduce a new \mgamc\ module, dubbed \maddecay, which
determines automatically the required final state multiplicity to
reach a given precision on the total width.
In addition, it generates the diagrams without any double counting,
estimates the contribution of each decay channel
and selects those that should be integrated numerically.
The inclusion of higher multiplicities may however still be insufficient in cases where
loop-induced decay modes are important or in cases where
threshold effect are large so that their resummation needs to be taken into account.

The paper is organized as follows: In Section~\ref{sec:fr} we give a short review of the \feynrules\ 
package and we introduce the new functions that can be used to compute two-body decay rates. In Section~\ref{sec:ufo} we present the extension of the UFO format to include the information on these two-body decays. In Section~\ref{sec:mg5} we describe the algorithm implemented in \maddecay\ to
generate all numerically relevant diagrams associated with $N$-body decay
channels and present how the code can be used
inside the \mgamc\ framework. Finally, in Section~\ref{sec:validation} we validate our implementations
by comparing partial widths for four selected models, namely the Standard Model, the Higgs Effective Field
Theory~\cite{Kniehl:1995tn,Shifman:1979eb}, the Strongly Interacting Light Higgs model~\cite{Giudice:2007fh}
and the Minimal Supersymmetric Standard Model.
Our conclusions are presented in Section~\ref{sec:conclusions}.

\section{Automatic computation of two-body decay rates with \feynrules}
\label{sec:fr}
\subsection{The \mathematica\ package \feynrules}

In this section we introduce one of the main actors involved in this work, the 
\feynrules\ package. We start by presenting a short 
review of the main functionalities of the package, and we focus on the automatic computation of decay rates within the \feynrules\ 
framework in Section \ref{sec:decay_fr}. 

In a nutshell, \feynrules\ is a \mathematica\ package that allows one for the automatic 
computation of the
Feynman rules of a quantum field theory model directly from its Lagrangian. 
It can be used with a large variety of physics models involving fields with spin of at
most two~\cite{Christensen:2009jx,Christensen:2013aua} and/or 
superfields~\cite{Duhr:2011se,Fuks:2012im}. The only requirements consist of Lorentz and 
gauge invariance. In other words, and more technically speaking, all indices appearing inside a Lagrangian must 
be correctly contracted. Apart from these restrictions, no further assumptions are made on 
the functional form of the Lagrangian, so that \feynrules\ can also be 
used to compute vertices associated with operators of dimension greater than four and in the context of any gauge choice.

In addition, \feynrules\ also contains several translation interfaces that allow one to output
the Feynman rules into a format readable by various Feynman diagram generators. Currently, 
dedicated interfaces exist for \calchep/\comphep~\cite{Pukhov:1999gg,Boos:2004kh,Pukhov:2004ca,%
Belyaev:2012qa}, \feynarts/\formcalc~\cite{Hahn:1998yk,Hahn:2000kx,Hahn:2006zy,Hahn:2009bf,%
Agrawal:2011tm},
\sherpa~\cite{Gleisberg:2003xi,Gleisberg:2008ta} and 
{\sc Whiz}-{\sc ard}/\oomega~\cite{Moretti:2001zz,Kilian:2007gr,Christensen:2010wz}. 
Furthermore, it is 
also possible to output the model information in the so-called \emph{Universal FeynRules Output}
(UFO) format, a format for the implementation of beyond the 
Standard Model theories into matrix element generators that is not tight to any existing 
code and that does not make any \textit{a priori} assumption on the structure of the 
interaction vertices that appear in the model. Finally, \feynrules\
also comes with specific computational modules that include, \eg, methods dedicated to
superspace calculations~\cite{Duhr:2011se}
or mass matrix diagonalization~\cite{Alloul:2013fw}.

The implementation of a model into \feynrules\ does not only require to
enter the Lagrangian,
but also the definitions of all particles and parameters that appear inside 
it. In other words, all symbols that are used to write the Lagrangian in 
\mathematica\ must be properly declared before Feynman rules may be calculated by 
\feynrules. The syntax for the declaration of the particles and parameters is
based on the original format for model files of \feynarts, extended by 
new options required by \feynrules. For example, the Standard Model $Z$-boson can be 
declared by including in the \feynrules\ model description,
\begin{verbatim}
  V[1] == { 
    ClassName       -> Z, 
    SelfConjugate   -> True,
    Mass            -> {MZ, 91.1876},
    Width           -> {WZ, 2.4952}
  }
\end{verbatim}
This declares a vector field ({\tt V[1]}) denoted by the symbol \texttt{Z} that is 
self-conjugate (\ie, that is equal to its own antiparticle) with a mass $M_Z$ of 91.187~GeV 
and a width $W_Z$ of 2.4952~GeV. At this stage, the numerical value of the width of the 
particles must be explicitly given when declaring a new instance of the 
particle class\footnote{If no numerical 
value is provided through the {\tt Width} attribute of the particle class, a default value 
of 1~GeV is assigned by \feynrules.}. The widths are, however, in general not independent input parameters of a model, but
related to other parameters of the theory, 
such as masses and coupling constants. Any consistent phenomenological analysis therefore requires the widths of all the particles appearing in a model to be re-evaluated every time the numerical values of the independent input parameters (consisting of so-called
`benchmark points') are changed. This is at odds with the fact that the widths are given as explicit numerical input parameters in a \feynrules\ model file.
  
The reason why explicit numerical values need to be specified in a \feynrules\ model lies in the interfacing to the matrix element generators. Indeed, as the only task of \feynrules\ is the computation of the tree-level Feynman rules from the Lagrangian, the widths, and even less so their numerical values, are not directly used at any stage by the code. However, when the Feynman rules are exported to one of the aforementioned matrix element generators, these codes often require the widths of the particles to be provided as numerical inputs, thus requiring the corresponding variables to be defined at the \feynrules\ level and the numerical values computed, \eg,
by making use of the generator of choice\footnote{Some tools offer
the possibility to compute the widths of all particles on the fly
when generating the matrix element associated with a given process. In
this case, including the numerical values of the widths at the \feynrules\
level is not mandatory.}.
Moreover, calculating all the branching ratios of the model particles
may also be required by tools such as parton shower programs that are
further interfaced to matrix element generators. 
This has to be repeated for each benchmark point under consideration,
which renders the entire approach highly inefficient as it involves a lot of redundant workload.

\subsection{Functions dedicated to the computation of decay widths}
\label{sec:decay_fr}
In this section, we describe the new functionalities of the \feynrules\ package related to the computation of all two-body decays of a model analytically. 
We start by giving a very brief review on two-body decays before presenting the new user functions implemented into \feynrules.

The leading-order decay rate of a heavy particle of mass $M$ into $N$ particles of mass $m_i$ is given by
\begin{equation}\label{eq:gamma}
\Gamma = \frac{1}{2|M|S}\int {\rm d} \Phi_N\,|\mathcal{M}|^2\ ,
\end{equation}
where $S$ denotes the phase space symmetry factor, $|\mathcal{M}|^2$ the averaged squared matrix
element and ${\rm d}\Phi_N$ the usual $N$-body phase-space measure in four dimensions
\begin{equation}\begin{split}\label{eq:PS_measure}
{\rm d}\Phi_N &\,= (2\pi)^4\,\delta^{(4)}\left(P-\sum_{i=1}^Np_i\right)\,\prod_{i=1}^N\frac{{\rm d}^4p_i}{(2\pi)^3}\,\delta_+(p_i^2-m_i^2)\\
&\,=(2\pi)^4\,\delta^{(4)}\left(P-\sum_{i=1}^Np_i\right)\,\prod_{i=1}^N\frac{{\rm d}^3p_i}{2\,(2\pi)^3\,E_i}\ .
\end{split}\end{equation}
In this expression $P=(M,\vec 0)$ stands for the four-momentum of the heavy decaying 
particle at rest and $p_i$ and $E_i$ are the momenta and energies of the decay products.
The absolute value included in Eq.~\eqref{eq:gamma}, rather unconventional, comes 
from the fact that in certain beyond the Standard Model theories involving Majorana fermions (\eg, the Minimal Supersymmetric Standard Model), it is possible to choose the phases of the fermion fields such that the mass is made negative. 

In the special case of a two-body decay, $N=2$, Lorentz invariance implies that the matrix element can only depend on the masses of the external particles, and we can write
\begin{equation}
\Gamma = \frac{\sqrt{\lambda(M^2,m_1^2,m_2^2)}\,|\mathcal{M}|^2}{16\,\pi\,S\,|M|^3}\,,
\end{equation}
where the K\"all\'en function reads
$\lambda(M^2,m_1^2,m_2^2)= (M^2-m_1^2-m_2^2)^2-4m_1^2m_2^2$.
The matrix element of a two-body decay only receives contributions from one single three-point vertex $\mathcal{V}$, and so it can be written as
\begin{equation}\label{eq:master}
|\mathcal{M}|^2 = \mathcal{V}_{\ell_1\ell_2\ell_3}^{a_1a_2a_3}\,\mathcal{P}^{\ell_1\ell'_1}_1\,\mathcal{P}^{\ell_2\ell'_2}_2\,\mathcal{P}^{\ell_3\ell'_3}_3\,(\mathcal{V}^*)_{\ell'_1\ell'_2\ell'_3}^{a_1a_2a_3}\,,
\end{equation}
where the color and spin indices of the particle $i$ are denoted by $\ell_i^{(')}$ and $a_i$. In addition, we have introduced the polarization tensor of the particle $i$, $\mathcal{P}_i$, which depends on its spin and its mass. 
As a consequence, the only dependence on the model is through the three-point vertex $\mathcal{V}$ computed by \feynrules, so that we have all the necessary ingredients to evaluate the two-body decay widths analytically. The reason why $N$-body decays with $N>2$ are not considered inside the \feynrules\ framework is due to the fact that,
on the one hand, 
the matrix element does no longer trivially decouple from the phase space and, on the other hand, the remaining phase-space integration might contain infrared divergences for massless particles in the final state. Moreover,
the double-counting arising from decay channels of different final-state
multiplicities requires a special treatment. The restriction to two-body decays provides in general a good estimate of the width of the particles. In some cases, it is however important to include 
at least three-body decay contributions. We will discuss how such cases are identified and
handled in the context of the \mgamc\ framework in Section~\ref{sec:mg5}.

Using the vertices provided by \feynrules, it is very easy to evaluate Eq.~\eqref{eq:master} analytically and to obtain the analytic results for \emph{all} two-body decay rates associated with any new physics model. 
In the rest of this section we describe the new functions included in the \feynrules\ package that allow one to perform this task on the example of the Standard Model 
implementation, which is included in the distribution of the package.

In order to compute the partial widths of the particles, it is necessary to first compute the vertices associated with the model in the usual way and to store them in some variable, \eg,
\begin{verbatim}
vertices = FeynmanRules[ LSM ];
\end{verbatim}
where {\tt LSM} denotes the variable containing the Standard Model Lagrangian. Once the vertices have been computed, we can immediately evaluate all two-body decays by issuing the command
\begin{verbatim}
decays = ComputeWidths[ vertices ];
\end{verbatim}
The function {\tt ComputeWidths[]} first selects all three-point vertices from the list {\tt vertices} that involve at least one massive particle and no ghost field and/or Goldstone boson. Next, the squared matrix elements are evaluated (in unitary gauge) using Eq.~\eqref{eq:master} and the results are stored in a list which contains entries of the form
\begin{equation*}
\textrm{{\tt\{\{}}\phi_1,\, \phi_2,\,\phi_3 \textrm{{\tt \}, }} \Gamma_{\phi_1\to\phi_2\,\phi_3} \textrm{{\tt \}}}\,.
\end{equation*}
First, we stress that the output of the function {\tt ComputeWidths[]} contains the analytic results for the decay rates for \emph{all} possible cyclic rotations of the external particles $\{\phi_1,\phi_2,\phi_3\}$ with a massive initial state, independently whether a given decay channel is kinematically allowed. The reason for this is that, while certain channels might be forbidden/allowed for a chosen set of numerical input parameters, this might not be the case for all possible choices of the external parameters. Second, the output of {\tt ComputeWidths[]} is also stored internally in the global variable \verb+FR$PartialWidth+. The use of this global variable will become clear below. Every time the function {\tt Com\-pu\-te\-De\-cays[]} is executed, the value of the global variable is overwritten, unless the option {\tt Save} of {\tt ComputeWidths[]} is set to {\tt False} (the default is {\tt True}) and in this case \verb+FR$PartialWidth+ remains unchanged.
Summations over possible internal gauge
indices in the analytic results are not performed, unless for indices related to the
fundamental and adjoint representations of $SU(3)$ in the case where the
user employs the conventional symbols for the representation matrices and
the corresponding index names presented in the \feynrules\
manual (see Section~6.1.5 of Ref.~\cite{Alloul:2013bka}).

While all the partial widths for all decay channels (kinematically allowed or not) are stored in the variables {\tt decays}  and \verb+FR$PartialWidth+,
\feynrules\ contains a set of functions that allows the user to read out directly certain entries of these lists. For example, issuing the command
\begin{quote}
{\tt PartialWidth[ \{}$\phi_1, \phi_2, \phi_3$ {\tt\}, decays ]}
\end{quote}
checks, based on the numerical values of the masses defined in the declaration of the particles, whether the decay $\phi_1\to\phi_2\,\phi_3$ is kinematically allowed, and if so,
returns the corresponding partial width $\Gamma_{\phi_1\to\phi_2\,\phi_3}$. The second argument of {\tt PartialWidth[]} is optional and could be omitted. If omitted, the partial widths stored in the global variable \verb+FR$PartialWidth+ are used by default. Similarly
\begin{quote}\flushleft
{\tt TotWidth[ }$\phi_1${\tt, decays ]};\\
{\tt BranchingRatio[ \{}$\phi_1, \phi_2, \phi_3${\tt\}, decays ]};
\end{quote} 
return the analytic results for the total decay rate and the branching ratio, whenever kinematically allowed.
Like for {\tt PartialWidth[]} the second argument is optional.
Furthermore, we stress that the analytic expressions can easily be evaluated numerically (using the numerical values for the masses and coupling constants defined in the model) with the built-in \feynrules\ function {\tt NumericalValue[]}.
Finally, it can be useful to update the information included in the original particle declarations by replacing the numerical value of the widths of all particles by the numerical values obtained with the function {\tt TotWidth}, which can be achieved by issuing the command
\begin{verbatim}
UpdateWidths[ decays ];
\end{verbatim}
where, as usual, the argument {\tt decays} is optional. After this command has been issued, the updated numerical results for the widths
are employed by the translation interfaces to matrix element generators.

\section{Extension of the UFO format to include decay information}
\label{sec:ufo}
In the previous section we have seen how \feynrules\ can be used to compute analytically all tree-level
two-body decays associated
with a beyond the Standard Model theory, how to update the numerical values for the total widths inside \feynrules\ and
how to export them to matrix element generators. This approach does however not yet solve the problem that widths are not independent parameters, and so they need to be re-evaluated for every benchmark point. While it is in principle possible to rerun \feynrules\ for each parameter set, this procedure is obviously highly inefficient. In the following, we present an extension of the UFO format~\cite{Degrande:2011ua} that
allows one to include the analytic results for the two-body decays.
Starting from \feynrules\ 2.0, the UFO interface automatically includes all two-body decays in the output. It is however possible to disable this feature by setting the option {\tt AddDecays} of the {\tt WriteUFO[]} function to {\tt False} (the default being  {\tt True}).

If included, information on the two-body decays is stored
into the UFO format in the file {\tt decays.py}. The content of this file contains declarations of instances of the class {\tt Decay} (defined in \verb+object_library.py+). Each instance of this class can be thought of as a collection of analytic formulas for the two-body partial widths of a given particle. For example, the two-body partial widths of the Higgs boson in the Standard Model are represented inside {\tt decays.py} as
\begin{quote}
{\tt Decay\textunderscore H = Decay(name = 'Decay\textunderscore H',}\\
\phantom{{\tt Decay\textunderscore }}{\tt particle = P.H,}\\
\phantom{{\tt Decay\textunderscore }}{\tt                partial\textunderscore widths = \{(P.W\textunderscore\textunderscore minus\textunderscore\textunderscore,P.W\textunderscore\textunderscore plus\textunderscore\textunderscore):'}$\Gamma_{H\to W^+\,W^-}${\tt',   }\\
\phantom{{\tt Decay\textunderscore partial\textunderscore widths = \{}}{\tt                                           (P.Z,P.Z):'}$\Gamma_{H\to Z\,Z}${\tt',}\\
\phantom{{\tt Decay\textunderscore partial\textunderscore widths = \{}}{\tt                                           (P.b,P.b\textunderscore\textunderscore tilde\textunderscore\textunderscore):'}$\Gamma_{H\to b\,\bar b}${\tt',}\\
\phantom{{\tt Decay\textunderscore partial\textunderscore widths = \{}}{\tt                                           (P.ta\textunderscore\textunderscore minus\textunderscore\textunderscore,P.ta\textunderscore\textunderscore plus\textunderscore\textunderscore):'}$\Gamma_{H\to \tau^+\, \tau^-}${\tt',}\\
\phantom{{\tt Decay\textunderscore partial\textunderscore widths = \{}}{\tt                                           (P.t,P.t\textunderscore\textunderscore tilde\textunderscore\textunderscore):'}$\Gamma_{H\to t\,\bar t}${\tt'}\\
\phantom{{\tt Decay\textunderscore }}{\tt                                )}
\end{quote}

where $\Gamma_{H\to ij}$ schematically represent the analytic formulas for the partial widths of the Higgs boson. The syntax used to write these analytic formulas in
\python\ form is identical to the syntax used in the UFO format for defining internal parameters, and we refer to Ref.~\cite{Degrande:2011ua} for details. Similar to the output of the {\tt ComputeWidths[]} function described in Section~\ref{sec:decay_fr}, \emph{all} possible decays are included, even if kinematically forbidden. In our
example\footnote{We consider the first two generations massless.},
this implies that the analytic formula for the decay of a Higgs boson into a top quark pair is also present
(even if not kinematically allowed for a light Higgs boson).
It is then up 
to matrix element generators to filter out at run time the kinematically allowed channels and to combine them consistently into the total width and branching ratios for a given particle.

The example of the Higgs is a case where tree-level two-body decays are not sufficient
for an accurate estimation of the width. One must indeed include important
contributions arising both from loop-induced diagrams and from three-body decays
via off-shell effects. While the inclusion of the loop-induced diagrams lies beyond
the scope of this paper, the next section will describe how three-body
decays can be handled with \mgamc.

\section{\maddecay\ -- a \mgamc\ module to compute decay widths}
\label{sec:mg5}

\mgamc\  is a suite of packages related to the computation of matrix elements \cite{Alwall:2011uj,Alwall:2008pm,%
Artoisenet:2012st,Artoisenet:2010cn,Hirschi:2011pa,Frederix:2009yq}. Two of its main usages consist of cross section
computation and event generation at the leading order (via the \madevent\ package~\cite{Maltoni:2002qb}) and next-to-leading
order (through the {\sc MC@NLO} framework~\cite{amcatnlo}) in perturbation theory.
In addition to their intrinsic accuracy, the precision of the results provided by the aforementioned packages
is limited by the precision of the calculations of the numerical values of the model parameters. In
particular, the way in which the widths of the particles have
been obtained plays a non-negligible role.
Although the total widths are in general estimated fairly well if only two-body decays are considered,
it is often necessary to include decays to higher-multiplicity final states.
In principle, we can generate all $N$-body tree-level decay processes directly in \mgamc\ and find the (partial) widths.\footnote{The same can be done with any Monte-Carlo generator with the same limitation.} However, in addition to the genuine $N$-body decay processes, this also generates cascade decays and radiation from ($n<N$)-body decays which can be accounted more efficiently from the parent lower-multiplicity decays. Also, often only a small subset of higher-multiplicity final states are relevant for the total width. Both factors hinder a balance between precision and efficiency using \mgamc\ alone.

\maddecay\ has been designed with the purpose to solve this issue. It is embedded as a module into
\mgamc, and allows the user to compute $N$-body partial decay widths for arbitrary values of $N$ at
tree-level. Assuming the narrow-width approximation\footnote{Even if those two assumptions are quite generic, there are often particles for which they are not satisfied, such as the Standard Model Higgs boson that has significant loop-induced decay modes or particles whose decay modes are threshold enhanced and where therefore resummation effects are important.}, 
the module iteratively builds diagrams from lower multiplicities while at the same time carefully removes contributions from cascade decays and radiative processes. We stress that \maddecay\ is not the first publicly available code to compute leading-order decay rates, but \maddecay\ has been
designed to go beyond what is done by existing tools like \calchep, {\sc Bridge}
or \herwig++\ in several aspects. First,
while \calchep\ can evaluate total decay widths on the fly, it only includes contributions
from three-body (four-body) decays only if there is no contribution from two-body
(three-body) decays.
Second, while \herwig++ and {\sc Bridge} are able to avoid contributions from cascade decays,
they are much more limited in the support of beyond the Standard Model theories than \maddecay,
which can handle, by its very design, any theory for which a UFO implementation exists.

In the remainder of this section we give a brief account of the algorithms underlying \maddecay. 
We first explain the algorithm for the diagram generation in Section~\ref{sec:algo}, followed by the algorithm to estimate numerical
relevance of a given channel compared to a total width and for the selection of the proper set of diagrams for event generation in Sections~\ref{sec:decay_estimator} and~\ref{sec:diagest}.
Finally, we introduce how to use the \maddecay\ module within the \mgamc\ framework in Section~\ref{sec:mg5_manuel}.

\subsection{The \maddecay\ algorithm}
\label{sec:algo}

The \maddecay\ module works in an iterative fashion. It begins by generating
all two-body decay diagrams, and
then iteratively adds extra final state particles to build up higher-multiplicity
diagrams\footnote{The program identifies stable particles in the very beginning and hence avoids the generation of their decay diagrams. 
The identification of the stable particles is based on the mass spectrum
and the particle interactions.}, until a certain maximal multiplicity $N_{\max}$ is reached. $N_{\max}$ is either provided as an input by the user or determined dynamically such that a requested numerical precision for the total decay width is reached (see Section~\ref{sec:mg5_manuel} for more details).
For $N$-body final states, all possible diagrams are first generated, including
the so-called contact diagrams that contain a single $N$-point interaction
vertex as well as diagrams derived from existing $n$-body decay processes (with $n<N$).
Diagrams corresponding to cascade decay and radiative processes are then removed following a procedure which is detailed in the remainder of this section.

\begin{figure}
\centering
 \begin{tabular}{cc}
 \includegraphics[scale=0.1]{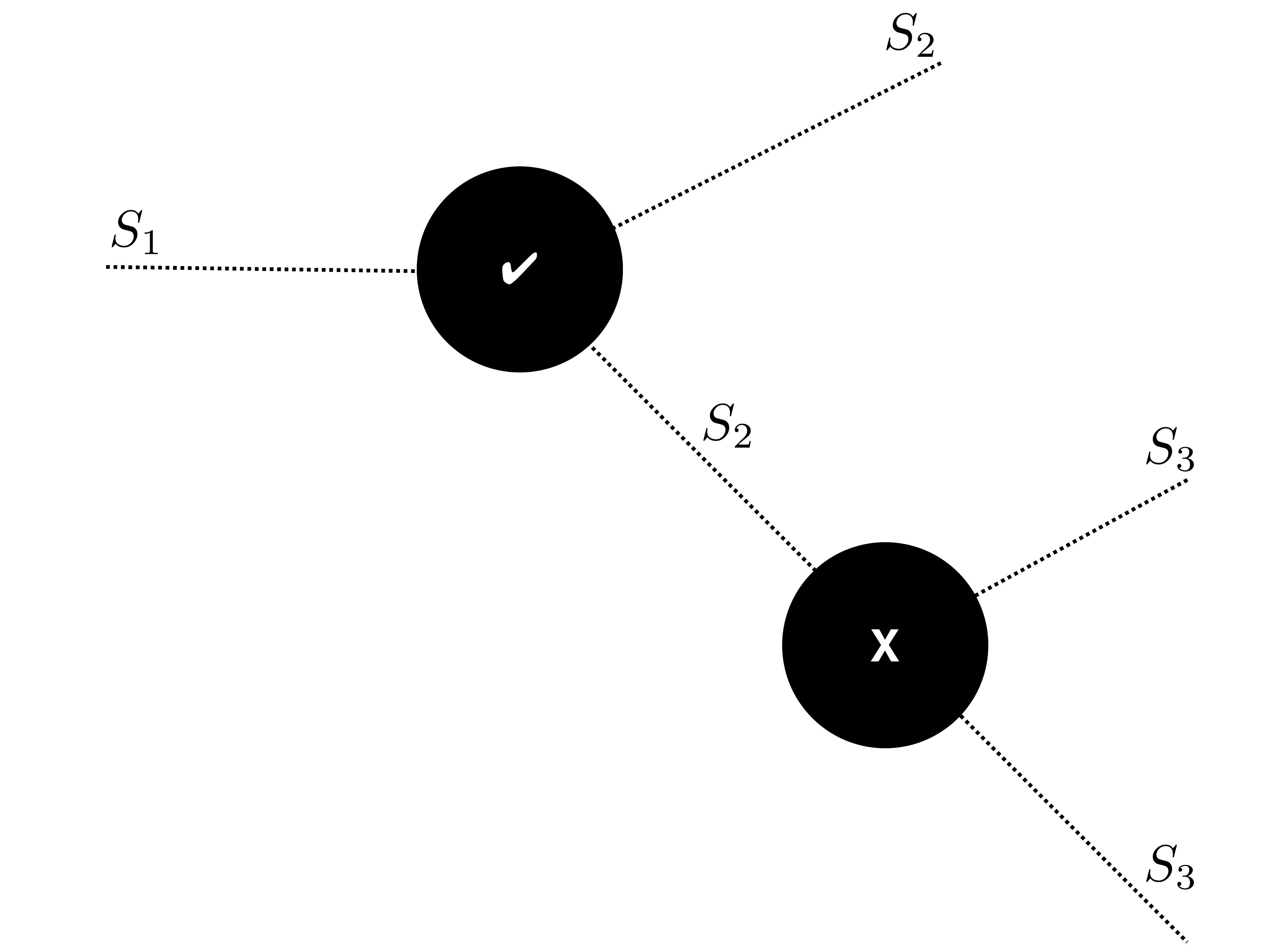}&
  \includegraphics[scale=0.1]{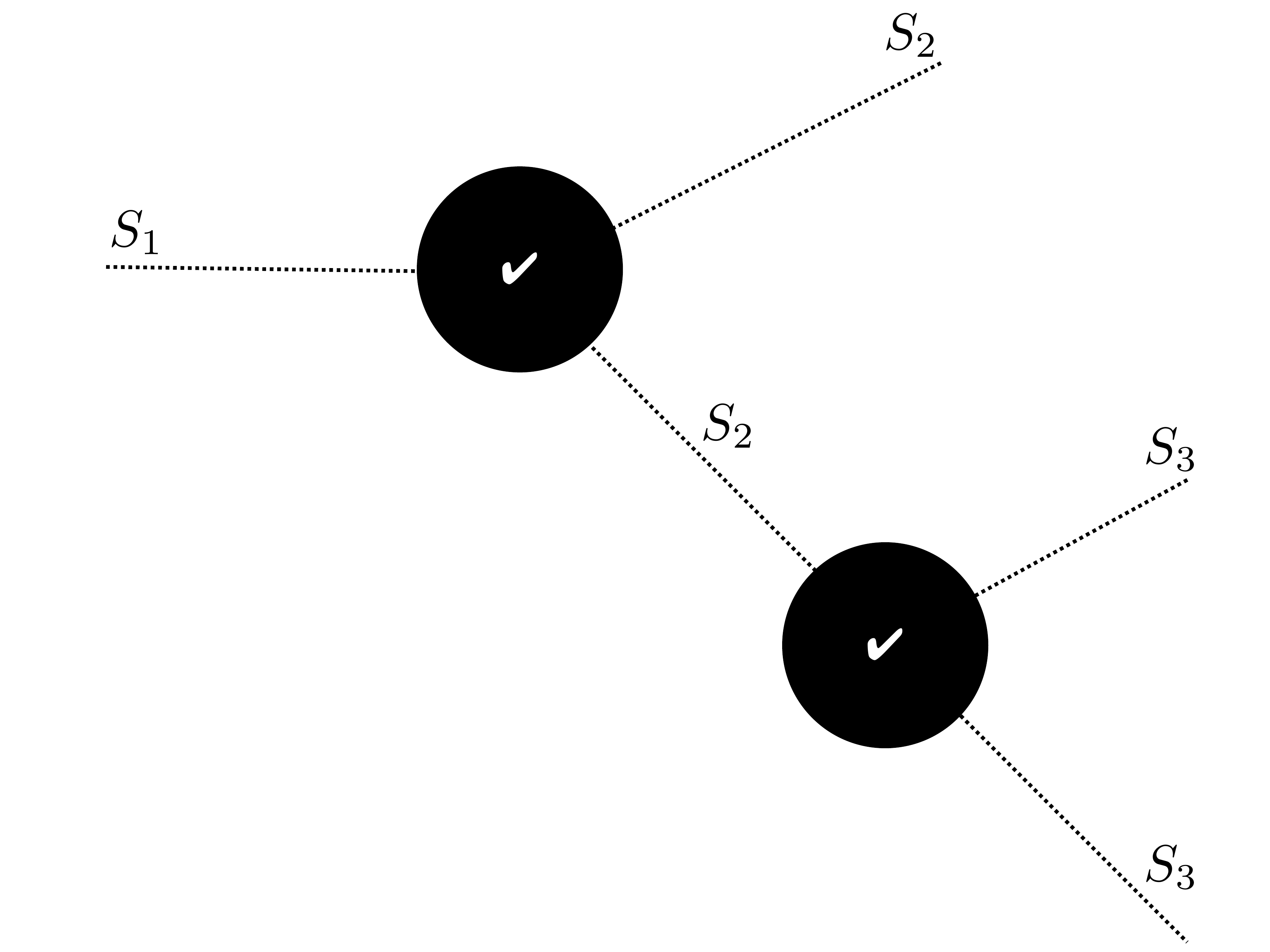}
    \includegraphics[scale=0.1]{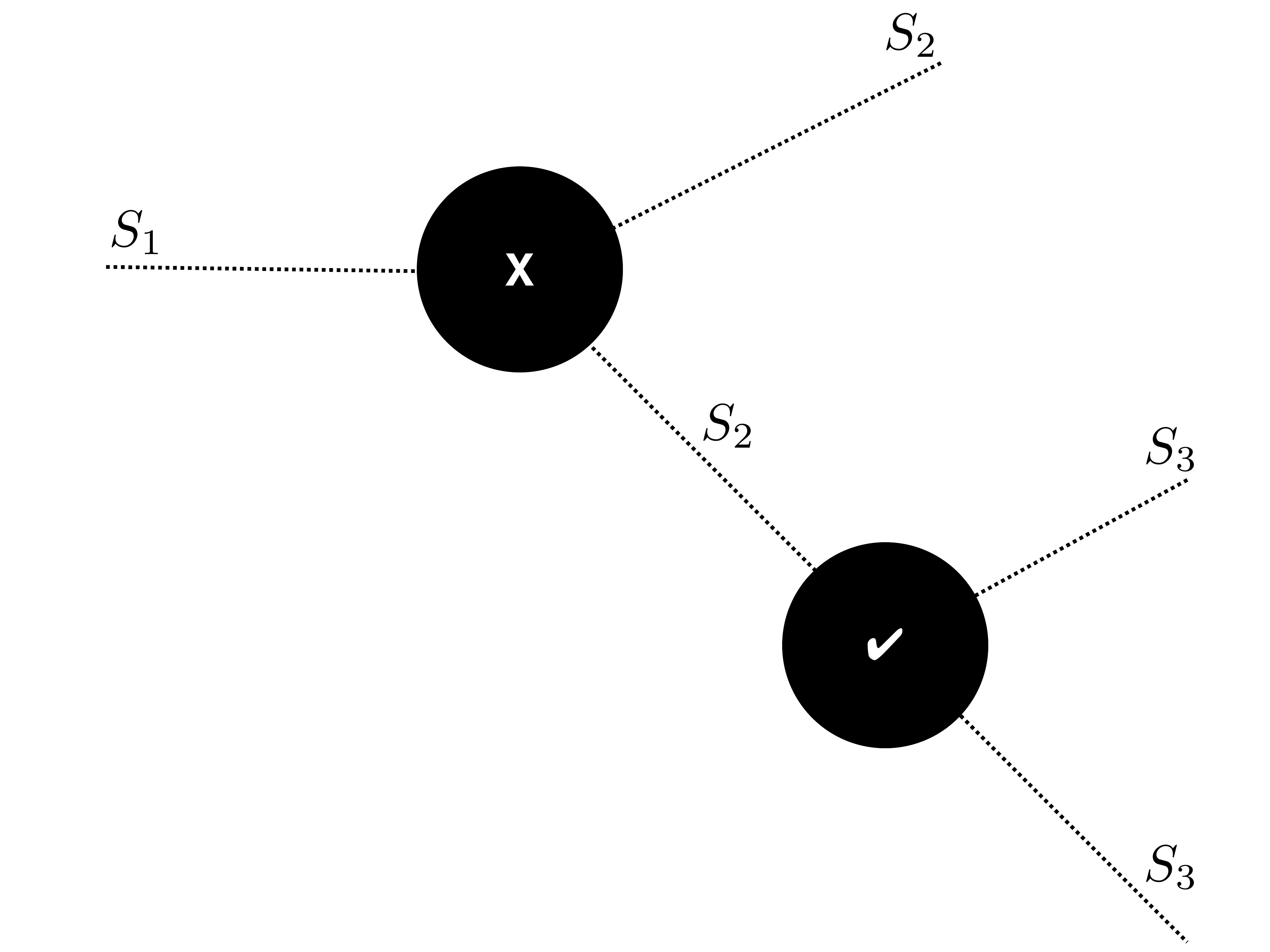}
  \end{tabular}
  \caption{\label{fig.lowerdecay}
Example of diagrams describing a schematical decay $S_1\to S_2 S_2^*\to S_2 S_3 S_3$. A cross (tick) on a vertex indicates that the associated decay is kinematically allowed (forbidden) with all three particles being on-shell. In the left-most diagram, the two-body decay  $S_1\to S_2 S_2$ is assumed to be kinematically viable, tagged as `open', while the process  $S_2\to S_3S_3$ is kinematically forbidden. Hence, this three-body decay mode of the $S_1$ particle must be mediated by an off-shell internal $S_2^*$ particle. \maddecay\ includes this mode as it cannot be captured by a cascade decay of an $S_2$ particle from a first $S_1\to S_2 S_2$ decay (see the central diagram), which would have been discarded.
In the right-most diagram, $S_1\to S_2 S_2$ is kinematically forbidden so that $S_2$ is forced to be off-shell. In this case, the three-body decay is possible and cannot be seen as a cascade of two-body decays, thus it is included by \maddecay.}
\end{figure}

In a first step, the algorithm removes diagrams corresponding to cascade decays.
Such diagrams contain at least one intermediate on-shell particle, and they are
generated by lower multiplicity processes with one or more subsequent
decays. 
Such diagrams need to be discarded as their contribution to the total width
is already accounted for by a lower-multiplicity decay.
For example, the decay $t\to b W^+ \to b \ell^+ \nu$ is a cascade decay. The
internal $W$-boson is indeed produced on-shell from the $t \to b W^+$ two-body
decay, and the $W$-boson further decays into a neutrino-lepton system. In contrast, diagrams
such as the one depicted in Fig.~\ref{fig.lowerdecay} are not defined as cascade
decay diagrams and their contribution is thus included into the width calculation.
In this case, although the $N$-body decay diagram contains an $n$-body decay
subdiagram with $n<N$ ($S_1 \to S_2\, S_2$ in the example of the
Fig.~\ref{fig.lowerdecay}), the $N$-body final state can only be produced if a 
internal particle ($S_2^*$ in Fig.~\ref{fig.lowerdecay}) is off-shell.

In order to remove cascade decay diagrams from the width calculation,
\maddecay\ tags all diagrams as `open'  or `closed' depending on whether or not they are 
kinematically
allowed, and only evaluates $N$-body diagrams that belong to one of the following
categories:
\begin{enumerate}
  \item The $N$-body decay diagram contains a closed $n$-body decay diagram with
    $n<N$. This configuration is the one of the most common higher-multiplicity
    decay diagrams.
  \item The $N$-body decay diagram contains an open $n$-body decay diagram with
    $n<N$ but the $N$-body final state can only be produced in the case where
    one of the internal particles is off-shell (as in
    Fig.~\ref{fig.lowerdecay}).
  \item The $N$-body decay diagram is a contact diagram (made from a single
    $N$-point interaction vertex).
\end{enumerate}
Each diagram is then tagged as `open' or `closed' and stored for the
generation of higher-multiplicity decays. Diagrams which do not belong to any of
the above categories are negligible in the narrow width approximation and are
omitted.

In a second step, the algorithm discards radiative diagrams. Indeed, as we perform a
tree-level (leading order) computation, we do not include any higher-order
contributions and an infrared-finite result can only be obtained after
coherently discarding all real-emission diagrams. We first investigate the
cascade decay topologies and discard any diagram that contains a particle
decaying into itself. This forbids, for example, the $ W^+\, \to\, W^+\,\gamma$,
$t \,\to\, t\, g$ and $ W^+\, \to \,W^+\, Z$ transitions. Additionally, we must check
that contact decay diagrams (made of a single $N$-point interaction vertex) do
not correspond to a radiative configuration. Equivalently, we need to
distinguish $N$-point interactions that emerge by imposing gauge invariance on
a $n$-point interaction vertex with $n<N$, as it arises when the derivatives
of a non-abelian gauge field are replaced by full field strength tensors, from
a leading-order new physics contribution. For example the dimension-six operator
$H^\dagger\, H\, G_\mu{}^\nu\, G_\nu{}^\mu$ gives rise to both $Hgg$ and $Hggg$
vertices. The second vertex is only necessary to render the first one gauge
invariant, and it does therefore not contribute to the width of the
Higgs boson into three gluons, because it is a higher-order QCD correction to the $H\,\to g\,g$ decay. In
contrast, if we consider the dimension-eight operator
$H^\dagger\, H\, G_\mu{}^\nu\, G_\nu{}^\rho\, G_\rho{}^\mu$, the leading-order
$h\,\to\,g\,g\,g$ contribution has to be taken into account, even at leading-order in QCD.
To disentangle those two cases, we generate all potential diagrams for given
initial and final state particles at a specific order in perturbation theory. If
a cascade decay topology is found, then the contact decay diagram is considered
as radiative and discarded. The definition of the perturbative order of a diagram
can be obtained from the UFO library (see Section~6.1.7 of
Ref.~\cite{Alloul:2013bka}).

\subsection{Fast estimation of $N$-body partial widths}
\label{sec:decay_estimator}

In order to avoid exporting all the `open'
$N$-body decay diagrams to \madevent\ for numerical integration, \maddecay\
performs a fast estimation of the contribution of each channel
assuming the absence of interference between diagrams contributing to the same final state,
and passes to \madevent\ only the numerically relevant decay modes\footnote{The interference is only neglected for the estimation of the width. All interference effects are of course correctly taken into account when computing the results for the partial widths.}.
The estimation is based on the formula
\begin{equation}
  \Gamma = \frac{1}{2 |M| S}\int {\rm d}\Phi_N \,|\mathcal{M}|^2
         \approx \frac{1}{2 |M| S} \, \textrm{LIPS}_N(M;m_1,\ldots,m_N)\, \langle |\mathcal{M}|^2 \rangle\ ,
\label{eq:estim}\end{equation}
where $\textrm{LIPS}_N(M;m_1,\ldots,m_N)$ is the Lorentz-invariant
$N$-body phase-space volume and $\langle |\mathcal{M}|^2 \rangle$ approximates the squared matrix element. In the following we describe in more detail how we evaluate the approximate decay rates.

\begin{figure}
\centering
\includegraphics[scale=0.2]{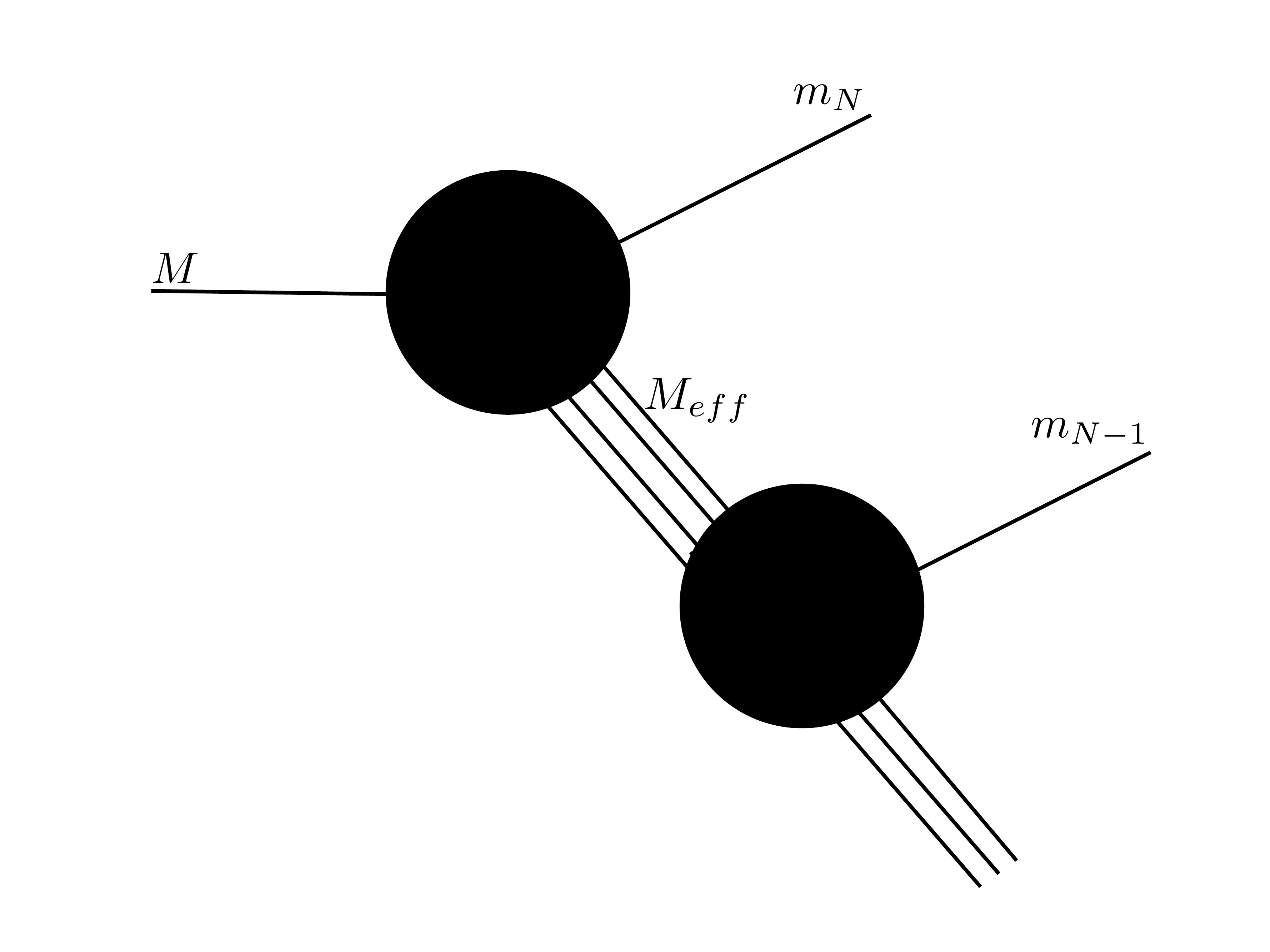}\\
\caption{\label{fig.ps_decomposition} Description of the recursive way
  of the computation of the phase-space volume factor in Eq.~\eqref{eq:PS_modified}. We denote
  by $M_{\rm eff}$ the effective mass associated with the other $N-1$ particles. See ref. \cite{Byckling:1969sx,James:1968gu} for more details.}
\end{figure}

We start by discussing the evaluation of the phase space volume.
Inserting 
\begin{equation}
1 = \int{\rm d}^4Q\,{\rm d} M_{\rm eff}^2\,\delta(Q^2-M_{\rm eff}^2)\,\delta^{(4)}\left(Q-\sum_{i=1}^{N-1}p_i\right)\,,
\end{equation}
into Eq.~\eqref{eq:PS_measure}, it is easy to see that (see Fig.~\ref{fig.ps_decomposition})
\begin{eqnarray}\label{eq:PS_modified}
&& \textrm{LIPS}_N(M;m_1,\ldots,m_N)\\
\nonumber    &&\qquad \  = \int^{M_{\rm max}^2}_{M_{\rm min}^2} \frac{\d M^2_{\rm eff}}{2\pi}\,
    \textrm{LIPS}_{2}(M;m_N,M_{\rm eff})\,\textrm{LIPS}_{N-1}(M_{\rm eff};m_1,\ldots,m_{N-1}) \ ,
\end{eqnarray}
where the two-body phase-space volume is given by
\begin{equation}
  \label{eq:PS_area_2body}
\textrm{LIPS}_2(M;m_1,m_2) =\frac{\sqrt{\lambda(M^2, m_1^2, m_2^2)}}{8\pi M^2}\,.
\end{equation}
Since the residual $(N-1)$-body phase-space measure is Lorentz invariant, we can
conveniently calculate it in the rest frame of
$P-p_N$. The mass of the effective mother particle (see
Fig.~\ref{fig.ps_decomposition}) is deduced from momentum conservation,
\begin{equation}
  M_{\rm eff}=\sqrt{M^2+m_N^2-2M\,E_N}\ ,
\end{equation}
where $E_N$ is the energy of
$N^{\rm th}$ particle in the rest frame of the mother particle.
This quantity $M_{\rm eff}$ ranges from
the production threshold of the remaining $N\!-\!1$ particles
(\mbox{$M_{\rm min}=\sum^{N-1}_{i=1}m_i$}) to a maximum value
$M_{\rm max}=M-m_N$ corresponding to the case where the $N^{\rm th}$
particle is at rest.
Finally, we approximate the exact expression of
Eq.~\eqref{eq:PS_modified} by
\begin{equation}\begin{split}
  \label{eq:PS_approximated}
   \textrm{LIPS}_N(M;m_1,\ldots,m_N)  &\,  \approx c_{ps}(M_{\rm max}^2-M_{\rm min}^2) 
    \frac{\sqrt{\lambda(M^2, m_N^2, M^2_{\rm mean})}}{16\pi^2 M^2}\\
    &\,\times
    \textrm{LIPS}_{N-1}(M_{\rm mean};m_1,\ldots,m_{N-1})\ ,
    \end{split}
\end{equation}
where the average effective mass is defined by
\mbox{$M_{\rm mean} \equiv (M_{\rm max}+M_{\rm min})/2$}.
Eq.~\eqref{eq:PS_approximated} consists of approximating the integral
by the area of a rectangle of height $M_{\rm mean}$.
As the integrand vanishes at both
$M_{\rm max}$ and $M_{\rm min}$
 and $\textrm{LIPS}_{N-1}(M_{\rm eff};m_1,\ldots,m_{N-1})$ vanishes
at the production threshold, the constant $c_{ps}=0.8$ has been
added to refine the estimation. This formula is recursive and the
recursion stops by using the analytic result
for the two-body phase space of Eq.~\eqref{eq:PS_area_2body}.

\setlength{\tabcolsep}{20pt}
\renewcommand{\arraystretch}{2.0}
\begin{table}
Table~\ref{table_propa_pol}: Simplified Feynman rules for propagators and polarization tensors
  for particles of different spins. We generically denote by $E$, $M$ and $\Gamma$
  the energy, mass and width of the particle under consideration. Concerning the width, the
algorithm either employs an estimation of the two-body decay width, if available,
or sets $\Gamma$ to zero, the subtraction of any resonant contribution allowing one to
avoid divergences (see the text for more details).
For massive particles, $f(E,M)=1+E^2/M^2$, and $f(E,M)=1$ for massless particles. Our implementation ignores massless propagator as those particles are stable.\\[.2cm]
\begin{center}
\begin{tabular}{c c c}
\hline \hline
Spin & Propagator & Polarization tensor \\
\hline \hline
  $0$           & $\displaystyle\frac{1}{(E^2-M^2+iM\Gamma)}$  & 1 \\
  $\frac{1}{2}$ & $\displaystyle\frac{E}{(E^2-M^2+iM\Gamma)}$  & $2E$\\
  $1$           & $\displaystyle\frac{1 - \frac{E^2}{M^2}}{(E^2-M^2+iM\Gamma)}$ & $f(E,M)$ \\
  $\frac{3}{2}$ & $\displaystyle\frac{\frac23 (E)(1-\frac{E^2}{M^2})}{(E^2-M^2+iM\Gamma)}$ & $2E\cdot f(E,M) $ \\
  $2$           &$\displaystyle \frac{\left(\frac76-\frac43\frac{E^2}{M^2}+\frac23\frac{E^4}{M^4}\right)}{(E^2-M^2+iM\Gamma)}$  &
    $f(E,M)^2 $ \\
\hline \hline
\end{tabular}
\textcolor{white}{\caption{\label{table_propa_pol}}}
\end{center}
\end{table}
\renewcommand{\arraystretch}{1}

The second ingredient to estimate the decay rate
is the average squared matrix element related to the decay process
under consideration. This quantity is approximated by
mimicking the calculation of standard Feynman diagrams using the
dedicated Feynman rules for the propagators and
external particles given in Table~\ref{table_propa_pol}.
The averaged matrix element square is then given by
\begin{equation}
  \label{eq:average_M}
\langle |\mathcal{M}|^2 \rangle=
\frac{N_{\textrm{color}}}{N_{\textrm{s}}} \times \left\vert  \prod_{\rm int}
  \textrm{Propa}(E)\right\vert ^2\,\times \prod_{\rm ext}\mathcal{P}(E) \,\times
\prod_{\rm vert} \left( \sum_i C_i \,\textrm{Lorentz}_i\right)^2\ .
\end{equation}

The first factor of this formula includes the average over the initial particle
spin states $1/N_{\textrm{s}}$ and the color multiplicity associated with the
diagram under consideration $N_{\textrm{color}}$. We recall that only color singlets,
triplets, sextets and octets are currently supported by \mgamc.

The second
factor in Eq.~\eqref{eq:average_M} describes the internal propagators
Propa$(E)$ appearing in the diagram, while the third factor includes
a product of the polarization tensors
$\mathcal{P}(E)$ of the external
particles. The Feynman rules of Table~\ref{table_propa_pol} have been chosen such as to
reproduce the correct results after a summation over all polarization and spin states.
Moreover, the available kinetic energy is assumed to be uniformly distributed
among the final-state particles, so that the energy $E_i$ of the $i^{\rm th}$ final-state
particle of mass $M_i$ reads
$E_i=(M - \sum_{j=1}^{N}M_j)/N+M_i$. The energy associated with an intermediate
propagator is then derived from the energy of the particles which it
decays into.

Finally, the last factor of Eq.~\eqref{eq:average_M} contains the interaction
vertices associated with the considered diagram. Each vertex is split up
into the different Lorentz structures $\textrm{Lorentz}_j$ that it contains, the
corresponding coupling constants being denoted by $C_j$. The Lorentz structures
are further simplified so that they can be evaluated very efficiently.
Each object that does not depend on the momenta (Dirac matrices,
chirality projectors, \etc) is replaced
by the identity while each object depending on the momenta ($p^\mu$, $\slashed{p}$,
\etc) is replaced by the energy of the relevant particle. We have checked that this treatment is
a good approximation also for non-trivial Lorentz structures (such as appearing in new physics models),
form factors as well as for derivative couplings.

\renewcommand{\arraystretch}{1.2}
\setlength{\tabcolsep}{10pt}
\begin{table}
 Table~\ref{table:fast_estimator}: Selection of partial decay widths
  of the heaviest sbottom in the context of the SPS1a MSSM
  scenario. We confront the estimations derived by \maddecay\ to the exact results.
  We only show partial widths larger than $10^{-7}$~GeV.\\[.2cm]
\begin{center}
\begin{tabular}{l c c c }
\hline \hline
 Process & Estimation of \maddecay  & Exact result\\
 & [GeV] & [GeV]\\
\hline
 $ \tilde b_2 \to \tilde \chi_1^-\, t\, h_1$ & 1.89e-03 & 1.04e-03 \\
 $ \tilde b_2 \to \tilde \chi_1^-\, t\, Z$ & 1.21e-03 & 1.27e-03 \\
 $ \tilde b_2 \to W^-\, t\, \tilde \chi_2^0$ & 9.78e-04 & 1.36e-03 \\
 $ \tilde b_2 \to \tilde \chi_2^-\, b\, W^+$ & 5.99e-04 & 1.88e-03 \\
 $ \tilde b_2 \to b\, Z\, \tilde\chi_2^0$ & 1.60e-04 & 2.88e-04 \\
 $ \tilde b_2 \to \bar \nu_\tau\, t\, \tilde \tau_2^-$ & 1.42e-04 & 3.36e-04 \\
 $ \tilde b_2 \to W^-\, t\, \tilde\chi_1^0$ & 1.34e-04 & 3.84e-04 \\
 $ \tilde b_2 \to \bar \nu_e\, t\, \tilde e_L^-$ & 1.31e-04 & 3.13e-04 \\
 $ \tilde b_2 \to \bar \nu_\mu\, t\, \tilde \mu_L^-$ & 1.31e-04 & 3.14e-04 \\
 $ \tilde b_2 \to W^-\, b\, \tilde\chi_1^+$ & 1.06e-04 & 2.04e-04 \\
 $ \tilde b_2 \to \tilde\nu_\tau\, t\, \tau^-$ & 7.52e-05 & 1.70e-04 \\
 $ \tilde b_2 \to \tilde\nu_e\, t\, e^-$ & 5.21e-05 & 1.23e-04 \\
 $ \tilde b_2 \to \tilde\nu_\mu\, t\, \mu^-$ & 5.21e-05 & 1.23e-04 \\
 $ \tilde b_2 \to b\, Z\, \tilde\chi_1^0$ & 2.20e-05 & 5.39e-05 \\
 $ \tilde b_2 \to b\, h^0\, \tilde\chi_2^0$ & 1.97e-05 & 1.33e-05 \\
 $ \tilde b_2 \to W^-\, b\, \tilde\chi_2^+$ & 8.53e-06 & 5.65e-06 \\
 $ \tilde b_2 \to b\, h^0\, \tilde\chi_1^0$ & 2.70e-06 & 2.19e-06 \\
 $ \tilde b_2 \to \bar b\, b\, \tilde b_1$ & 1.51e-06 & 8.58e-07 \\
 $ \tilde b_2 \to \bar \nu_\tau\, t\, \tilde \tau_1^-$ & 7.77e-07 & 2.09e-06 \\
 $ \tilde b_2 \to b\, Z\, \tilde \chi_4^0$ & 6.86e-07 & 3.54e-07 \\
 $ \tilde b_2 \to \tilde{b}_1^*\, b\, b$ & 6.37e-07 & 2.04e-07 \\
 $ \tilde b_2 \to b\, Z\, \tilde\chi_3^0$ & 5.34e-07 & 3.05e-07 \\
 $ \tilde b_2 \to \bar u\, u\, \tilde b_1$ & 2.24e-07 & 3.48e-07 \\
 $ \tilde b_2 \to \bar c\, c\, \tilde b_1$ & 2.24e-07 & 3.48e-07 \\
 $ \tilde b_2 \to \bar s\, s\, \tilde b_1$ & 1.77e-07 & 4.50e-07 \\
 $ \tilde b_2 \to \bar d\, d\, \tilde b_1$ & 1.77e-07 & 4.50e-07 \\
 $ \tilde b_2 \to \mu^+\, \mu^-\, \tilde b_1$ & 1.01e-07 & 1.02e-07 \\
 $ \tilde b_2 \to e^+\, e^-\, \tilde b_1$ & 1.01e-07 & 1.02e-07 \\
 $ \tilde b_2 \to \bar \nu_e\, \nu_e\, \tilde b_1$ & 5.39e-08 & 2.03e-07 \\
 $ \tilde b_2 \to \bar \nu_\mu\, \nu_\mu\, \tilde b_1$ & 5.39e-08 & 2.03e-07 \\
 $ \tilde b_2 \to \bar \nu_\tau\, \nu_\tau\, \tilde b_1$ & 5.39e-08 & 2.03e-07 \\
 \hline \hline
\end{tabular}
\textcolor{white}{\caption{\label{table:fast_estimator}}}
\end{center}
\end{table}
\renewcommand{\arraystretch}{1.}

The decay diagram estimation method has been validated in the framework of various models,
and especially in the Standard Model and the Minimal Supersymmetric Standard Model (MSSM).
The validation procedure has shown that the estimator can be safely used to select which decay channels
are relevant to be computed numerically. Representative
examples of the validation tests are presented in Table~\ref{table:fast_estimator}, where we compare the
estimation of all partial widths of the heaviest bottom squark $\tilde b_2$ to the corresponding
exact result derived from a {\sc Bridge}-\madevent-\calchep\ comparison
in the context of the SPS1a MSSM benchmark scenario~\cite{Allanach:2002nj}.
One can observe that the estimations reproduce, in all cases, the correct order of magnitude
for each channel.

\subsection{Estimation of the numerically relevant Feynman diagrams}
\label{sec:diagest}

In Section~\ref{sec:decay_estimator}, we have presented a method
allowing \maddecay\ to estimate whether a given decay
channel is numerically relevant once the corresponding
Feynman diagrams have been generated. In this section, we present a second routine
dedicated to a rough and quick estimate of the partial width
associated with any decay process
prior to diagram generation. This allows \maddecay\
to proceed with diagram generation only in cases where the considered decay mode can yield
a non-negligible contribution to the total width.

This new method relies on the derivation of the partial width related to
a kinematically allowed $(N\!+\!1)$-body decay by combining a $N$-body decay
with the two-body decay of one of the $N$ final-state
particles. The average squared matrix element
$\langle |\mathcal{M}|^2 \rangle_{N+1}$
is hence derived from the knowledge of the average squared matrix elements
$\langle |\mathcal{M}|^2 \rangle_N$ and $\langle |\mathcal{M}|^2 \rangle_i$
that respectively describe the $1\to N$ decay of the mother particle (of mass $M$)
and the $1\to 2$ decay of the $i^{\rm th}$ final-state particle (of mass $M_i$).
Calculating $\langle |\mathcal{M}|^2 \rangle_N$ as in
Eq.~\eqref{eq:average_M} after approximately fixing the energy of each final-state
particle to $M/(N+1)$, the squared matrix element
$\langle |\mathcal{M}|^2 \rangle_{N+1}$ can be written as \footnote{
As an order of magnitude estimate, we take $\vert\textrm{Propa}(E_i)\vert^2 \approx 1/(0.5 M^2)^2$ for every propagator in $\langle |\mathcal{M}|^2 \rangle_{N}$. This also avoids the propagators to be accidentally on-shell.}
\begin{equation}
  \label{eq:M_new}
\langle |\mathcal{M}|^2 \rangle_{N+1}=
\langle |\mathcal{M}|^2 \rangle_{N} \times \langle |\mathcal{M}|^2
\rangle_{i}\times
\frac {\vert\textrm{Propa}(E_i)\vert^2}{\mathcal{P}_i(M/(N+1)) \mathcal{P}_i(M_i)}\ ,
\end{equation}
where one considers one specific value of $i$ (the summation will be
performed below).
The last factor of the above formula shows that the external polarization
tensors $\mathcal{P}^2_i(M/(N+1))$ and $\mathcal{P}^2_i(M_i)$ of the $i^{\rm th}$ particle
that are included in the matrix elements are replaced
by a propagator $\vert\textrm{Propa}(E_i)\vert^2$. In addition, the color
multiplicity and spin average factor have been neglected.

Since the decay of the $i^{\rm th}$ particle is a two-body decay, the related
matrix element is related to the associated partial width as
\begin{equation}
  \label{eq:M_2}
  \Gamma_i = \frac{1}{2|M_i|}\textrm{LIPS}_2(M_i) \times 
    \langle |\mathcal{M}|^2 \rangle_{i} \approx
  \frac{1}{16 \pi |M_i|} \times 
    \langle |\mathcal{M}|^2 \rangle_{i}\ ,
\end{equation}
where $\textrm{LIPS}_2(M_i) \approx 1/(8\pi)$ is the two-body phase-space volume given in
Eq.~\eqref{eq:PS_area_2body} evaluated by neglecting, for simplicity,
all final-state particle masses.

In order to get an estimate of the $N\!+\! 1$-body decay partial width, it is
also necessary to estimate the $N\!+\!1$-body phase space volume given by
Eq.~\eqref{eq:PS_approximated}. To this aim, we naively assume that
$(M_{\rm max}^2-M_{\rm min}^2)
\approx (M/2)^2$ and that $\sqrt{\lambda(M^2, m_N^2, M^2_{\rm mean})} \approx M^2$.
Both of these approximations give the correct orders of magnitude and allow one,
by iterating, to evaluate the phase space volume as
\begin{equation}
  \label{eq:PS_Nadd1}
  \textrm{LIPS}_{N+1}(M) \approx c_{ps}\times
  \left(\frac{M}{8\pi}\right)^2\textrm{LIPS}_N(M) \\
   \sim \left(c_{ps} \left(\frac{M}{8\pi}\right)^2\right)^{N-2}\times \frac{1}{8\pi}
\end{equation}
The estimation of the $1\to N\!+\!1$ partial width is then given by
combining Eq.~\eqref{eq:M_new}, Eq.~\eqref{eq:M_2}, and Eq.~\eqref{eq:PS_Nadd1},
\begin{equation}
\Gamma_{N+1} \approx \widetilde{\Gamma}_N \times \sum_{i} c_{ps}\Gamma_i \frac{M_i M^2}{4 \pi}
   \frac {\vert\textrm{Propa}(E_i)\vert^2}{\mathcal{P}_i(M_0/(N+1)) \mathcal{P}_i(M_i)},
\end{equation}
where $\widetilde{\Gamma}_N$ is obtained from the phase space volume computed as in
Eq.~\eqref{eq:PS_Nadd1} and $\langle |\mathcal{M}|^2 \rangle_N$ calculated
as described previously. Additionally, the summation over $i$ includes
all possible decays of any of the $N$ final-state particles of the $1\to N$
process.

\renewcommand{\arraystretch}{1.22}
\setlength{\tabcolsep}{30pt}
\begin{table}
 Table~\ref{tab:valid_fastest_diag}: Relative contribution of the three-body decays to the total width of all
  massive particles in the context of the SPS1a MSSM scenario. We approximate the
  total width by neglecting any decay to four particles or more and
  confront the estimated results of \maddecay\ (third column) to the exact results (second column). The partial widths of three-body decays are conservatively constrained by our analytical estimation.
  This shows that the order of magnitude is either correctly evaluated or over-estimated which is conservative.\\[.2cm]
\begin{center}
\begin{tabular}{l c c}
\hline \hline
Particle & $\Gamma_3/(\Gamma_3+\Gamma_2)$ & Estimation \\
\hline\hline
$ t$ & 7.06e-09 & 4.62e-06 \\
$ Z$ & 5.16e-07 & 3.46e-04 \\
$ W^+$ & 0.00e+00 & 0.00e+00 \\
$ h^0$ & 6.76e-02 & 1.82e+00 \\
$ H^0$ & 1.55e-03 & 2.42e-03 \\
$ A^0$ & 1.30e-03 & 2.13e-03 \\
$ H^+$ & 2.53e-03 & 3.99e-03 \\
$ \tilde d_L$ & 8.19e-03 & 5.92e-03 \\
$ \tilde d_R$ & 1.44e-04 & 9.35e-02 \\
$ \tilde s_L$ & 8.15e-03 & 5.92e-03 \\
$ \tilde s_R$ & 1.44e-04 & 9.35e-02 \\
$ \tilde b_1$ & 5.08e-03 & 1.21e-03 \\
$ \tilde b_2$ & 1.07e-02 & 3.83e-02 \\
$ \tilde u_L$ & 9.06e-03 & 5.46e-03 \\
$ \tilde u_R$ & 1.45e-04 & 2.38e-02 \\
$ \tilde c_L$ & 9.08e-03 & 5.46e-03 \\
$ \tilde c_R$ & 1.44e-04 & 2.38e-02 \\
$ \tilde t_1$ & 2.28e-03 & 1.89e-03 \\
$ \tilde t_2$ & 4.74e-03 & 4.47e-04 \\
$ \tilde e_L^-$ & 3.77e-05 & 2.65e-04 \\
$ \tilde e_R^-$ & 1.63e-11 & 2.08e-06 \\
$ \tilde\mu_L^-$ & 3.80e-05 & 2.65e-04 \\
$ \tilde\mu_R^-$ & 1.65e-11 & 2.08e-06 \\
$ \tilde\tau_1^-$ & 0.00e+00 & 0.00e+00 \\
$ \tilde\tau_2^-$ & 5.99e-04 & 4.14e-04 \\
$ \tilde\nu_e$ & 1.26e-08 & 1.01e-04 \\
$ \tilde\nu_\mu$ & 1.27e-08 & 1.01e-04 \\
$ \tilde\nu_\tau$ & 3.76e-04 & 2.44e-04 \\
$ \tilde g$ & 4.94e-04 & 1.12e-02 \\
$ \tilde\chi_2^0$ & 5.08e-03 & 9.18e-02 \\
$ \tilde\chi_3^0$ & 1.16e-03 & 9.88e-03 \\
$ \tilde\chi_4^0$ & 1.34e-03 & 1.20e-01 \\
$ \tilde\chi_1^+$ & 8.09e-03 & 7.78e-01 \\
$ \tilde\chi_2^+$ & 1.45e-03 & 2.93e-02 \\
 \hline \hline
\end{tabular}
\textcolor{white}{\caption{\label{tab:valid_fastest_diag}}}
\end{center}
\end{table}
\renewcommand{\arraystretch}{1.}

This method to estimate whether a given process is relevant has been
carefully validated in the context of many new physics theories. For illustrative
purposes, we focus on the SPS1a MSSM scenario~\cite{Allanach:2002nj}
and present in Table~\ref{tab:valid_fastest_diag}
the relative contribution of the three-body decay modes to the total width
(computed by neglecting any decay to four particles or more) of all massive particles
of the model.
We compare the exact results to those estimated by \maddecay. This shows that the order
of magnitude is either correctly evaluated or over-estimated, which is conservative
since this implies that the diagrams related to a non-negligible channel are always
generated.

\subsection{The \mgamc\ interface to \maddecay}
\label{sec:mg5_manuel}

The \maddecay\ module is fully embedded into the \mgamc\ framework and
there are currently two ways to use it. Either \maddecay\ is directly
called from a \mgamc\ or \madevent\ shell, or it is instead run
on the fly, at the time of event generation or cross section
computation. In all situations, using \maddecay\
requires  a valid UFO model. In
the cases where the file \texttt{decays.py} is available (see Section~\ref{sec:ufo}),
the analytic results for two-body decays are directly used by \maddecay.
Otherwise, all partial widths are computed numerically.

The first way to use \maddecay\ is to call it from a \mgamc\
or \madevent\ command interface,
respectively initiated via the commands {\tt ./bin/mg5\textunderscore aMC} and {\tt ./bin/madevent}.
This method is in particular useful for creating a valid \texttt{param\_card.dat}. 
Width calculations are performed by issuing the
command
\begin{verbatim}
  compute_widths PARTICLE_NAME [OTHER PARTICLE] [OPTIONS]
\end{verbatim}
where \texttt{PARTICLE\_NAME} refers either the name of a particle or its associated Particle Data Group
(PDG) code. The user can enter more than one particle name or PDG code, and can also
use the keyword \texttt{all} to calculate the width of all the particles\footnote{It is
recommended to compute as many widths as possible in one single execution of the command to reduce the overhead.}.
The following options are allowed:
\begin{itemize}
\item {\tt --body\_decay=value} [default: 4.0025]. The code ignores
  $N$-body decay contributions when either $N$ is larger than the integer part of
  \texttt{value} or when the estimated error for the total width is lower than the decimal part of
  \texttt{value}. In the case where the integer/decimal part of \texttt{value} is set to zero,
  the associated condition is ignored. For instance,
\begin{itemize}
   \item[$~$] {\tt --body\_decay=3} enforces the computation of all two- and three-body decay channels.
   \item[$~$] {\tt --body\_decay=0.01} stops width computations when the estimated error on the total width is lower than 1\%.
   \item[$~$] {\tt --body\_decay=3.01} stops a width computation either when all three-body decay contributions have
    been included or when the estimated error on the total width is lower than 1\%.
\end{itemize}
\item {\tt --min\_br=value} [default: 0.000625]. If the estimation of the branching ratio associated with a given
  decay mode is found below \texttt{value}, the channel is not integrated numerically and the
  mode will not appear in the decay table. If not specified explicitly, \texttt{value} is set
  to the decimal part of the \texttt{body\_decay} parameter divided by four.
\item {\tt --precision\_channel=value} [default:  0.01]. Required relative precision on each individual channel
  when integrated numerically.
\item {\tt --path=value} [default value for \mgamc: path to the UFO model] [default value for {\sc MadEvent}: \texttt{./Cards/param\_card.dat}].
  The path to the {\tt param\_card.dat} file to use during the numerical evaluation.
\item {\tt --output=value} [default: overwrite input file]. The path where to store the new {\tt param\_card.dat} file
  that includes the computed widths.
\end{itemize} 

The second way to use \maddecay\ is to run it on the fly either through a \madevent\ or a {\sc aMC@NLO} session.
Both programs start by checking the \texttt{param\_card.dat} file. If any of the widths is set to the value \texttt{Auto}
then \maddecay\ is called on the fly (with the default options) to evaluate these quantities.
The \texttt{param\_card.dat} file is then overwritten before any further computation.
For instance, if the {\tt param\_card.dat} file provided to \madevent\ contains the lines
\begin{verbatim}
DECAY   6 Auto # WT
DECAY  23 Auto # WZ
DECAY  24 2.047600e+00 # WW
DECAY  25 5.753088e-03 # WH
\end{verbatim}
\madevent\ then calls \maddecay\ to compute the total width and decay table of the top quark and the $Z$-boson,
the widths of the $W$-boson and Higgs boson remaining unchanged. The original {\tt param\_card}.dat file
is subsequently overwritten, the above lines being replaced by
\begin{verbatim}
#      PDG        Width
DECAY  6   1.491472e+00
#  BR             NDA  ID1    ID2   ...
   1.000000e+00   2    24  5 # 1.49147214391
#      PDG        Width
DECAY  23   2.441755e+00
#  BR             NDA  ID1    ID2   ...
   1.523651e-01   2    3  -3 # 0.372038381506
   1.523651e-01   2    1  -1 # 0.372038381506
   1.507430e-01   2    5  -5 # 0.368077510282
   1.188151e-01   2    4  -4 # 0.290117391009
   1.188151e-01   2    2  -2 # 0.290117391009
   6.793735e-02   2    16  -16 # 0.165886384843
   6.793735e-02   2    14  -14 # 0.165886384843
   6.793735e-02   2    12  -12 # 0.165886384843
   3.438731e-02   2    13  -13 # 0.0839653943458
   3.438731e-02   2    11  -11 # 0.0839653943458
   3.430994e-02   2    15  -15 # 0.0837764784469
#
#      PDG        Width
DECAY  24   2.047600e+00
#
#      PDG        Width
DECAY  25   5.753088e-03
\end{verbatim}
where the corresponding partial widths are given under the form of a
comment, at the end of each line.
Since the code returns not only the total widths, but also all partial widths, the output file
is perfectly suitable to be passed to a parton shower program that can then further decay the
unstable particles possibly present in the hard events.

\section{Illustrative examples} \label{sec:validation}
As examples of usage of the tools presented in the previous sections,
we focus on two-body  and three-body partial width computations and compare,
for a set of specific new physics theories, results provided by \feynrules,
\maddecay, several public tools and analytic formulas available in the literature.

\subsection{Two-body decays}
In this section, we focus on two-body decay widths and perform
various calculations in
the framework of the Standard Model, the Strongly Interacting Light Higgs
(SILH) model~\cite{Giudice:2007fh} and the SPS1a MSSM benchmark
scenario~\cite{Allanach:2002nj}. We first
compute for each of these models
all two-body partial widths with \feynrules\ and then
numerically compare the results to those returned by the
\maddecay\ module of \mgamc.
Moreover, in the case of the MSSM, we also confront the
\feynrules\ results to the analytic formulas of Ref.~\cite{Bozzi:2007me}.
Agreement has been found in all cases, which validates our implementations
in particular for theories involving higher dimensional operators
(\textit{cf}.\ the SILH model)
and those with Majorana fermions (\textit{cf}.\ the MSSM).
A selection of numerical results can be found in
Table~\ref{SMtable}, Table~\ref{SILHtable} and Table~\ref{MSSMtable}
for the Standard Model, the SILH model and the MSSM, respectively.

\setlength{\tabcolsep}{20pt}
\renewcommand{\arraystretch}{1.22}
\begin{table}
  Table~\ref{SMtable}: Selection of partial decay widths in the
  framework of the Standard Model, as computed by \feynrules\ and \maddecay.\\[.2cm]
  \begin{center}
  \begin{tabular}{l c r}
\hline  \hline
Decay mode & \feynrules\ [GeV] & \maddecay~[GeV]\\ 
\hline
\hline
$h \to b \, \bar{b}$ & $0.005390$  & $0.005391$ \\ 
$h \to \tau \, \bar{\tau}$ & $0.0002587$ & $0.0002587$ \\ 
$h \to c \, \bar{c}$ & $0.0003967$ &  $0.0003967$\\
\hline 
$W^+ \to e^+ \, \nu_e $ & $0.2225$ & $ 0.2225$ \\
$W^+ \to \tau^+ \, \nu_{\tau} $ & $0.2223$ & $ 0.2224$ \\
$W^+  \to u \, \bar{d}$ & $0.6336$ & $ 0.6336$ \\
$W^+  \to c \, \bar{s}$ & $0.6333$ & $0.6334$ \\
$W^+  \to c \, \bar{d}$ & $0.03401$ & $0.03402$ \\
$W^+  \to u \, \bar{s}$ & $0.03403$  & $ 0.03403$ \\
\hline
$Z \to e^- \,  e^+$ & $0.08329$ & $0.08329$ \\ 
$Z \to \tau^- \, \tau^+$ & $0.0831$ & $ 0.0831$  \\
$Z \to \nu_{e} \, \bar{\nu}_{e}$ & $0.1658$ & $ 0.1659$ \\
$Z \to u \, \bar{u}$ & $0.2841$ & $0.2842$ \\ 
$Z \to d \, \bar{d}$ & $0.3667$ & $0.3667$ \\
$Z \to c \, \bar{c}$ & $0.2838$ & $ 0.2839$ \\
$Z \to b \, \bar{b}$ & $0.3627$ & $ 0.3628$ \\
\hline
$t \to b \, W^+$ & $1.466$ & $1.467$ \\
\hline \hline
\end{tabular}
\textcolor{white}{\caption{\label{SMtable}}}
\end{center}
\end{table}
\renewcommand{\arraystretch}{1.}

\setlength{\tabcolsep}{20pt}
\renewcommand{\arraystretch}{1.22}
\begin{table}
  Table~\ref{SILHtable}: Higgs boson partial decay widths in the
  framework of the SILH model, as computed by \feynrules\ and \maddecay.\\[.2cm]
  \begin{center}
  \begin{tabular}{l c r}
\hline \hline
Decay mode & \feynrules\ [GeV] & \maddecay~[GeV]\\ 
\hline
\hline
$h \to \gamma \gamma$ & $6.447 \, \mbox{e-10}$  & $6.447 \, \mbox{e-10}$ \\ 
$h \to g\, g$ & $ 7.523 \, \mbox{e-06}$  &   $ 7.524 \,\mbox{e-06}$ \\ 
$h \to \gamma \, Z $ &  $4.026 \, \mbox{e-11}$  & $4.026 \, \mbox{e-11}$ \\
\hline \hline
\end{tabular}
\textcolor{white}{\caption{\label{SILHtable}}}
\end{center}
\end{table}
\renewcommand{\arraystretch}{1.}

\setlength{\tabcolsep}{20pt}
\renewcommand{\arraystretch}{1.22}
\begin{table}
  Table~\ref{MSSMtable}: Selection of partial decay widths in the
  framework of the SPS1a MSSM scenario, as computed by \feynrules\ and \maddecay.\\[.2cm]
\begin{center}
  \begin{tabular}{l c r}
\hline \hline
 Decay mode & \feynrules\ [GeV] & \maddecay~[GeV]\\ 
\hline
\hline
$\tilde{\chi}^0_4 \to \tilde{\chi}^+_1 \, W^-$ & $0.6451 $  & $0.6451$ \\ 
$ \tilde{\chi}^0_4 \to \tilde{\chi}^0_1 \, Z$ & $0.05567 $  &   $ 0.05568$ \\ 
$ \tilde{\chi}^+_2 \to \tilde{\chi}^0_1 \, W^+$ & $0.1682 $  &   $0.1683$ \\
$ \tilde{\chi}^+_2 \to \tilde{\chi}^+_1 \, Z$ & $0.5755$  &   $0.5756 $ \\
$ \tilde{u}_6 \to \tilde{u}_3 \, Z$ & $1.39 $  &   $1.4$ \\[1ex]
\hline \hline
\end{tabular}
\textcolor{white}{\caption{\label{MSSMtable}}}
\end{center}
\end{table}
\renewcommand{\arraystretch}{1.}

\subsection{Three-body decays}

In this section, we address the calculation of tree-body decay widths and compare results obtained
with \maddecay\ to those available in the literature. We first consider the
Higgs Effective Field Theory (HEFT)~\cite{Kniehl:1995tn,Shifman:1979eb} where the Standard Model Lagrangian
is supplemented by an additional dimension-six operator allowing the Higgs boson to directly couple to gluons
and photons with coupling strengths tuned to reproduce the corresponding loop-induced vertices of the
Standard Model\footnote{We stress that only these dimension-six operators are included, and not the full set of such operators.}. Next, as a second example, we focus on the MSSM, which allows us to further test
the programs in the case of processes with Majorana particles.

\setlength{\tabcolsep}{20pt}
\renewcommand{\arraystretch}{1.22}
\begin{table}
  Table~\ref{table:higgs_decay}: Higgs boson partial decay widths in the
  framework of the HEFT model, as computed by \maddecay\
  and by {\sc SMCalc}.\\[.2cm]
\begin{center}
  \begin{tabular}{l c r}
\hline  \hline
Decay mode: & \maddecay\ [GeV] &  {\sc SMCalc}~[GeV]\\ 
\hline
$h \to b \, \bar{b}$ & $ 0.00430 $  & $ 0.00430 $ \\ 
$h \to c \bar c$ & $0.000496$ & $0.000496$\\
$h \to \tau \, \bar{\tau}$ & $0.000259 $ & $ 0.000259 $  \\ 
$h \to g \, g $ & $ 0.000195 $ & $0.000195 $ \\ 
$h \to W\,  W^{*}\to W\, f\, f $ & $ 0.000771 $ & $0.000775$  \\ 
$h \to Z\,  Z^*  \to Z\, f\, f $  & $8.44\mbox{e-05}$ & $ 8.40\mbox{e-05} $ \\
$h \to \gamma \, \gamma $ & $ 9.70\mbox{e-06} $ &  $ 9.731\mbox{e-06}$\\
\hline 
\end{tabular}
\textcolor{white}{\caption{\label{table:higgs_decay}}}
\end{center}
\end{table}
\renewcommand{\arraystretch}{1.}

In Table~\ref{table:higgs_decay}, we present results for Higgs boson decays
in the framework of the HEFT model and compare partial widths calculated
by \maddecay\ to results returned by {\sc SMCalc}\footnote{{\sc SMCalc} is a program that contains analytical formul\ae\ for all leading-order SM partial widths. It is available from
Ref.~\cite{SMcalc}.}. For the sake of completeness, we have
included both two-body and three-body decay channels
and found good agreement between the two programs.

\setlength{\tabcolsep}{3.5pt}
\renewcommand{\arraystretch}{1.22}
\begin{table}
  Table~\ref{table:maddecay_mssm}: Selection of total decay widths in the
  framework of the SPS1a MSSM scenario, as computed from the \texttt{decay.py} file
  generated by \feynrules~(first column), \maddecay~with
  default option values (second column) and by enforcing three-body decays (third
  column) and by {\sc Bridge} (fourth column).\\[.2cm]
\begin{center}
  \begin{tabular}{c c c c c}
\hline \hline
 Particle & \feynrules  & \maddecay & \maddecay  & {\sc Bridge} \\
 & Two-body [GeV]  & Default~[GeV] & Three-body~[GeV] & Three-body~[GeV] \\ 
\hline
$ \tilde \chi_1^+$ & 1.704e-02   & 1.718e-02  &  1.718e-02 & 1.724e-02 \\
$ \tilde \chi_2^+$ & 2.487       & 2.488         &  2.488     & 2.485\\
$ H^+ $    & 6.788e-01  & 6.788e-01  &  6.802e-01 & 6.780e-01\\
$ \tilde b_1 $    & 3.736 & 3.736 & 3.740 & 3.731\\
$ \tilde b_2 $    & 8.016e-01  & 8.071e-01  & 8.094e-01  & 8.100e-01\\
$ \tilde \chi^0_2 $    & 2.078e-02  & 2.087e-02  & 2.088e-02  & 2.082e-02\\
$ \tilde \chi^0_3 $    & 1.916e+00 & 1.916e+00 & 1.916e+00 & 1.914e+00 \\
\hline \hline
\end{tabular}
\textcolor{white}{\caption{\label{table:maddecay_mssm}}}
\end{center}
\end{table}
\renewcommand{\arraystretch}{1.}

In Table \ref{table:maddecay_mssm}, we consider the SPS1a MSSM scenario
and compare the results obtained with the \texttt{decay.py}
file generated by \feynrules\ (that only includes two-body decay modes),
with \maddecay\ when using all the default option values described in Section~\ref{sec:mg5_manuel},
with \maddecay\ by enforcing the calculation of all three-body decay channels and with
{\sc Bridge} (when including all three-body decay modes).
Agreement below the percent level has been found. Additionally, this shows that for the
SPS1a MSSM scenario, three-body decay contributions to the total widths
are negligible in a very good approximation.

\setlength{\tabcolsep}{6pt}
\renewcommand{\arraystretch}{1.22}
\begin{table}
  Table~\ref{table:maddecay_speed}: Time necessary to compute all particle total widths
  in the context of the HEFT model and the SPS1a MSSM scenario by making use of 
  the \texttt{decay.py} file generated by \feynrules\ only (first column), 
  \maddecay\ with all default option values and {\sc Bridge} by restricting the computation
  to two-body decays only (second column) and after including three-body decays too (fourth column).
  For the tests, we have employed a machine with a 2.3 GHz Intel Core i7 processor and 16 GB of
  memory (1600 MHz DDR3).\\[2cm]
\begin{center}
\begin{tabular}{l c c c c}
\hline \hline
Model & \feynrules\  & {\sc Bridge } & \maddecay\  & {\sc Bridge } \\ 
 & Two-body  & Two-body & Default & Three-body \\ 
\hline
\hline
HEFT model & 0.6 s & 60s & 40s & 114 s \\
SPS1a MSSM scenario & 12 s & 13min 43s  & 84 s & 1h47 \\
\hline \hline
\end{tabular}
\textcolor{white}{\caption{\label{table:maddecay_speed}}}
\end{center}
\end{table}
\renewcommand{\arraystretch}{1.}

Finally, we compare the speed of \maddecay\ (using the \texttt{decay.py}
file generated by \feynrules) to the one of {\sc Bridge}
and indicate in Table~\ref{table:maddecay_speed} the time necessary
to compute the total widths of all model particles in the context of both the HEFT model and
the SPS1a MSSM scenario. When only the analytic formulas for the two-body decays (implemented
in the \texttt{decay.py} file generated by \feynrules) are used
(first column of the table), \maddecay\ turns out to be much faster than {\sc Bridge} (second
column of the table) as no diagram generation has been required at all for the first case.
The time necessary to generate the \texttt{decay.py} UFO file has however
not been included (a few seconds and
minutes for the HEFT model and MSSM, respectively) as this has to be performed only once for each model.
When including three-body decay contributions, one can note the formidable gain in time when using
\maddecay\ instead of {\sc Bridge} (last two columns of the table) thanks to
the usage of the fast
estimator of \maddecay\ which allows one to only compute numerically relevant
diagrams.

\section{Conclusion}
\label{sec:conclusions}
In this paper, we have presented new routines of the \feynrules\ package dedicated
to the computation of two-body partial widths, so that the latter can now
be calculated automatically and analytically
from the knowledge of the Lagrangian alone. The
UFO format has been extended accordingly to include
all the relevant information.
This extension is currently supported by \mgamc, which uses it
to calculate particle widths at run-time. In addition, a new module
of \mgamc, named \maddecay, has been developed with the aim of 
computing $N$-body decay widths in full generality and in an efficient manner
(possibly at run-time when using the matrix-element generator).
The \maddecay\ routines automatically remove all
the subprocesses that are numerically negligible, tune the final-state multiplicity to reach a given target precision on the total widths and avoid the double-counting of any
channel. All the computations are done at tree level/leading order and rely on the narrow width approximation. As such, 
\maddecay\ cannot be used to obtain reliable predictions for the widths in cases where higher-order effect are important.
Our codes have been carefully validated against
existing older programs and published results in the literature
in the case of the Standard Model,
Higgs Effective Field Theories, the Strongly Interacting
Light Higgs model and the Minimal Supersymmetric Standard Model.
The widths obtained by \maddecay\ are accurate enough to be used in any LO Monte-Carlo generator as long as the 
narrow width approximation holds. 

\section*{Acknowledgments}
We want to thank C. Degrande and F. Maltoni for useful discussions on this project
and D.~Goncalves Netto and K.~Mawatari for their constructive bug reports.
O.M. want to thank the IPPP center for its hospitality during the time of this project.
B.F.\ was supported in part by the French ANR 12 JS05 002 01 BATS@LHC and by
the Theory-LHC-France initiative of the CNRS/IN2P3. CHS is supported
by the DOE under grant number DE-SC0010255. DGO is supported by Theoretische Forschungen auf dem Gebiet der Elementarteilchen (SNF).
OM is `Chercheur scientifique logistique postdoctoral F.R.S.-FNRS', Belgium.
This work was partly supported by the Research Executive Agency (REA) of the European Union under the Grant Agreement number PITN-GA-2010-264564 (LHCPhenoNet), by MCnetITN FP7 Marie Curie Initial Training Network PITN-GA-2012-315877, by the IISN
``MadGraph'' convention 4.4511.10, by the IISN ``Fundamental interactions''
convention 4.4517.08, and in part by the Belgian Federal Science Policy Office
through the Interuniversity Attraction Pole P7/37. 

\bibliography{biblio}

\end{document}